\documentstyle[preprint,aps,psfig]{revtex}
\newcommand{\beq}{\begin{equation}}
\newcommand{\eeq}{\end{equation}} 
\newcommand{\beqa}{\begin{eqnarray}}
\newcommand{\eeqa}{\end{eqnarray}} 
\newcommand{\ba}{\begin{array}} 
\newcommand{\ea}{\end{array}} 

\begin{document}
\draft

\widetext 
\title{Parametric Resonance Phenomena in Bose-Einstein Condensates: 
Enhanced Quantum Tunneling of Coherent Matter Pulses} 
\author{Luca Salasnich} 
\address{Istituto Nazionale per la Fisica della Materia, 
Unit\`a di Milano, \\ 
Dipartimento di Fisica, Universit\`a di Milano, \\ 
Via Celoria 16, 20133 Milano, Italy \\
E-mail: salasnich@mi.infm.it} 
\maketitle 
\begin{abstract} 
We investigate the quantum tunneling of a Bose-Einstein condensate 
confined in a optical trap. 
We show that periodic pulses of coherent matter are emitted from the trap 
by using an oscillating energy barrier. Moreover, 
the emitted fraction of condensed atoms  
strongly increases if the period of oscillation of 
the height of the energy barrier 
is in parametric resonance with the period of oscillation of 
the center of mass of the condensate inside the potential well. 
Our model is analyzed by numerically solving the nonpolynomial 
Schr\"odinger equation (NPSE), an effective one-dimensional 
equation which describes the macroscopic wavefunction 
of Bose condensates under transverse harmonic confinement. 
The range of validity of NPSE is discussed and compared with 
that of Gross-Pitaevskii equation. 
\end{abstract} 

\vskip 1. truecm

\narrowtext

\newpage 

\section{Introduction} 

Nowadays dynamical properties and resonant phenomena of Bose-Einstein 
condensates are the subject of many experimental 
and theoretical invertigations. 
Recently, the parametric excitation of cold trapped atoms 
in far-off-resonance optical lattices 
has been experimentally obtained by modulating the potential depth [1,2]. 
Another kind of parametric resonance, namely 
the parametric resonance between the collective oscillations 
of a trapped Bose-Einstein condensate 
and the oscillations of the confining harmonic trapping potential 
has been demonstrated with numerical simulations [3]. 
Moreover, by imposing particular values of anisotropy 
to the trapping harmonic potential, it has been shown that 
different collective modes of the condensate are in resonance 
and the oscillations can be chaotic [4,5]. 
\par 
In this paper we discuss a mechanism to generate 
coherent matter pulses by means of macroscopic quantum tunneling  
of a Bose condensate through the barrier of a potential well [6]. 
In our model the Bose-Einstein condensate is confined in the vertical 
axial direction by two Gaussian optical barriers and in the transverse 
direction by a magnetic or optical harmonic potential. 
The zero-temperature dynamics of Bose-Einstein condensate 
under transverse harmonic confinement is well described 
by the one-dimensional nonpolynomial Schrodinger equation (NPSE) we have 
recently derived from the three-dimensional Gross-Pitaevskii 
equation (3D GPE) [7,8]. 
In the first part of the paper we discuss the range of validity 
of the 3D GPE and that of the NPSE. In the second part of the paper 
by the numerical integration of NPSE we show that  
it is possible to generate periodic waves of 
coherent matter by periodically changing the height of the lower-lying 
Gaussian barrier. Moreover, the emission probability is strongly enhanced 
when the oscillation of the height of the energy barrier is 
in parametric resonance with the oscillation of the center of mass of the 
condensate inside the potential well. 

\section{Conditions on the Gross-Pitaevskii Equation} 

The time-dependent three-dimensional Gross-Pitaevskii equation (3D GPE) [9], 
that describes the macroscopic wavefunction (or order parameter)  
$\psi({\bf r},t)$ of the Bose condensate, is given by 
\beq 
i\hbar {\partial \over \partial t}\psi ({\bf r},t)= 
\left[ -{\hbar^2\over 2m} \nabla^2 
+ U({\bf r}) + g N |\psi ({\bf r},t)|^2 \right] \psi({\bf r},t)  \; , 
\eeq  
where $U({\bf r})$ is the external trapping potential and 
$g={4\pi \hbar^2 a_s/m}$ is the scattering amplitude and $a_s$ 
the s-wave scattering length. $N$ is the number of condensed 
bosons and the wavefunction is normalized to one. 
Note that the 3D GPE is a nonlinear Schr\"odinger (Hartree) equation 
and the nonlinear term is due to the interatomic 
interaction described by the two-body pseudo-potential 
$V({\bf r},{\bf r}')= g \delta^{(3)}({\bf r}-{\bf r}')$ [10]. 
\par 
The 3D GPE is accurate to describe a condensate 
of dilute gas of bosons only near zero temperature, 
where thermal excitations can be neglected [10]. 
The dilute gas condition is given by 
$a_s^3 \rho \ll 1$, where $a_s$ is the s-wave scattering length 
of the interatomic potential and $\rho$ is the density of the 
Bose gas. This diluteness condition means that the range ($a_s$) 
of the interatomic potential is much smaller than the average 
distance ($\rho^{-1/3}$) between atoms. 
By introducing the healing length 
\beq 
\xi = {1\over \sqrt{8\pi a_s \rho}} \; , 
\eeq 
the dilute gas condition can be written as $a_s \ll \xi$. 
Note that the condensate can be dilute but strongly 
interacting if its interatomic energy ($4\pi \hbar^2 a_s \rho/m$) 
is much larger than its kinetic energy. 
In the case of an axially symmetric harmonic confinement 
by using the healing length $\xi$ the dilute and strongly-interacting 
gas condition can be written as $a_s \ll \xi \ll a_H$, 
where $a_H=\sqrt{\hbar\over m\omega_H}$ is the characteristic 
length of the harmonic confinement 
with $\omega_H = (\omega_{\bot} \omega_z)^{1/3}$. 
\par 
Another interesting condition is the one-dimensional gas condition. 
One has the 1D regime when the potential energy of the 
transverse confinement is much larger than the interatomic 
energy of the Bose condensate. In the case of a transverse 
harmonic confinement, again using the healing length $\xi$, 
the 1D gas condition for a Bose condensate 
can be written as $a_{\bot} \ll \xi$, where 
$a_{\bot}=\sqrt{\hbar\over m\omega_{\bot}}$ is the characteristic 
length of the transverse harmonic confinement. It follows that, 
if the inequalities $a_s \ll a_H < a_{\bot} \ll \xi$ 
are satisfied, then the Bose condensate is dilute, 
cigar-shaped and one-dimensional. 
\par 
It is important to observe that for $\xi \gg a_{1D}$, where 
$a_{1D}$ is the 1D scattering length given by $a_{1D}=a_{\bot}^2/a_s$ [10], 
the Bose gas enters in the Tonks-Girardeau regime: a one-dimensional and very dilute 
gas of impenetrable Bosons for which the Bose-Einstein 
condensation is absent [11]. 
Thus in the Tonks-Girardeau regime the 3D GPE cannot be applied. 
In conclusion, the 3D GPE is valid if the healing length $\xi$ of the 
Bose condensate satisfies the condition 
\beq 
a_s \ll \xi < a_{\bot}^2/a_s \; ,  
\eeq 
and it describes a one-dimensional Bose condensate if 
$a_s \ll a_{\bot} \ll \xi < a_{\bot}^2/a_s$. 

\section{Nonpolynomial Schr\"odinger Equation} 

In many experiments with Bose-Einstein condensates the external trapping 
potential can be described by a harmonic potential 
in the transverse direction 
and a generic potential in the axial direction: 
\beq 
U({\bf r})={1\over 2}m\omega_{\bot}^2(x^2+y^2) + V(z) \; . 
\eeq
This external potential suggests to map the 3D GPE into an effective 
1D equation, which simplifies greatly the solution of the 3D GPE. 
This problem is not trivial due to the nonlinearity 
of the GPE. 
\par 
The 3D GPE is the Euler-Lagrange equation 
of the following Lagrangian density 
\beq
{\it L} = \psi^*({\bf r},t) \left[ i\hbar {\partial \over \partial t} 
+ {\hbar^2 \over 2 m} \nabla^2 - U({\bf r}) 
-{1\over 2}gN |\psi ({\bf r},t)|^2 \right] \psi({\bf r},t)  \; . 
\eeq 
For the wavefunction we choose the following variational ansatz 
\beq 
\psi({\bf r},t) = \phi(x,y,t;\sigma(z,t)) \; f(z,t) \; , 
\eeq 
where both $\phi$ and $f$ are normalized to one and $\phi$ is represented 
by a Gaussian:  
\beq
\phi(x,y,t;\sigma(z,t)) = { e^{-(x^2+y^2)\over 2 \sigma(z,t)^2} 
\over \pi^{1/2} \sigma(z,t)} \; . 
\eeq
\par 
Moreover we assume that the transverse wavefunction $\phi$ is slowly 
varying along the axial direction with respect 
to the transverse direction, 
i.e. $\nabla^2 \phi \simeq \nabla_{\bot}^2 \phi$  
where $\nabla_{\bot}^2={\partial^2 \over \partial x^2}+  
{\partial^2 \over \partial y^2}$.  
By inserting the trial wave-function in the Lagrangian density  
and after spatial integration along $x$ and $y$ variables  
the Lagrangian density becomes  
\beq
{\it L} = f^* \left[ i\hbar 
{\partial \over \partial t}  + {\hbar^2\over 2 m} 
{\partial^2\over \partial z^2} - V -  {1\over 2} 
g N {\sigma^{-2}\over 2\pi} |f|^2 
- {\hbar^2 \over 2m}\sigma^{-2} - 
{m\omega_{\bot}^2\over 2} \sigma^2 \right] f \; .  
\eeq 
The Euler-Lagrange equations with respect  
to $f^*$ and $\sigma$ read  
\beq
i\hbar {\partial \over \partial t}f=  
\left[ -{\hbar^2\over 2m} {\partial^2\over \partial z^2}  
+ V + g N {\sigma^{-2}\over 2\pi} |f|^2   
+ \left({\hbar^2 \over 2m}\sigma^{-2} 
+ {m\omega_{\bot}^2\over 2} \sigma^2  \right) \right] f  \; ,  
\eeq   
\beq {\hbar^2 \over 2m}\sigma^{-3} - 
{1\over 2}m\omega_{\bot}^2 \sigma  + 
{1\over 2} g N {\sigma^{-3}\over 2\pi} |f|^2 = 0 \; .  
\eeq 
The second Euler-Lagrange equation reduces to an algebric  
relation providing a one to one correspondence between $\sigma$  
and $f$: $\sigma^2 = a_{\bot}^2 \sqrt{1 + 2 a_s N |f|^2}$,  
where $a_{\bot}=\sqrt{\hbar \over m\omega_{\bot}}$ 
is the oscillator  length in the transverse direction. 
One sees that $\sigma$  depends implicitly on $z$ and $t$ 
because of the  space and time dependence of $|f|^2$.  
Inserting this result in the first equation one finally obtains  
$$
i\hbar {\partial \over \partial t}f=  
\left[ -{\hbar^2\over 2m} {\partial^2\over \partial z^2}  
+ V + {g N \over 2\pi a_{\bot}^2} {|f|^2\over 
\sqrt{1+ 2a_sN|f|^2} }  \right.  
$$ 
\beq 
\left.  
+ {\hbar \omega_{\bot}\over 2}   
\left( {1\over \sqrt{1+ 2 a_sN|f|^2} } 
+ \sqrt{1+ 2a_sN|f|^2} \right) \right] f  \; . 
\eeq  
This equation is a time-dependent non-polynomial 
Schrodinger equation (NPSE). 
\par 
We observe that under the condition $a_sN|f|^2 \ll 1$ 
one has $\sigma^2=a_{\bot}^2$ and NPSE reduces to 
\beq 
i\hbar {\partial \over \partial t}f= 
\left[ -{\hbar^2\over 2m} {\partial^2\over \partial z^2} 
+ V + {g N \over 2\pi a_{\bot}^2} |f|^2 \right] f  \; ,  
\eeq 
where the additive constant $\hbar \omega_{\bot}$ 
has been omitted because it does not affect the dynamics. 
This equation is a 
1D Gross-Pitaevskii equation (1D GPE). The nonlinear coefficient $g'$ 
of this 1D GPE can be thus obtained from the nonlinear coefficient 
$g$ of the 3D GPE by setting $g'=g/(2\pi a_{\bot}^2)$ 
such that $V(z,z')=g'\delta(z-z')=(2\hbar^2/ma_{1D})\delta(z-z')$ 
is the effective 1D interatomic pseudo-potential 
and the effective 1D scattering length is $a_{1D}=a_{\bot}^2/a_s$, 
in agreement with the definition of the previous section. 
Note that the limit $a_s N|f|^2 \ll 1$ is precisely the condition 
for the one-dimensional regime where the healing length $\xi$ 
is larger than $a_{\bot}$. But, as shown in the previous section, 
the 3D GPE and consequently the NPSE are no more valid if 
$\xi \gg a_{\bot}^2/a_s$. It follows that the 1D GPE is valid only 
where the condition $a_s^2/a_{\bot}^2 < a_sN|f|^2 \ll 1$ 
is satisfied. 
\par 
In papers [9,10] we have tested the accuracy of the NPSE 
in the determination of the ground-state and collective 
oscillations of the condensate with axial harmonic 
confinement and also in the description 
of tunneling through a Gaussian barrier. 
In particular, we have compared NPSE with the the full 3D GPE. 
The conclusion of these investigations 
is that NPSE is very accurate in the description 
of cigar-shaped condensates ($a_z/a_{\bot} \geq \sqrt{10}$) 
both in the 3D regime and in the 1D regime 
($a_s\ll\xi< a_{\bot}^2/a_s$). 
\par
Note that the NPSE has been recently used by Massignan 
and Modugno to study 
the dynamics of a Bose condensate in a optical lattice [12]. 

\section{Periodic emission of matter waves} 

In this section we use the NPSE to study a 
mechanism which produces periodic emission of 
matter waves by means of the quantum tunneling 
of a Bose condensate. We consider a cigar-shaped 
condensate confined by a harmonic potential in the 
transverse direction and under the action of 
the gravity potential $m{\it g}z$ and two Gaussian functions 
that model a confining potential well in the vertical 
axial direction: 
\beq 
V(z)= V_1 \; e^{-(z-z_1)^2/ \sigma^2} + 
V_2 \; e^{-(z-z_2)^2/ \sigma^2} + m {\it g} z \; . 
\eeq 
Such a configuration can be experimentally obtained by using  
two blue-detuned laser beams (perpendicular to 
the axial direction) which are modelled by the two Gaussian 
potentials. By varying the intensity of the lower-lying 
laser beam one controls the height of the energy barrier 
and therefore the tunneling probability. 
The periodic height given by: 
\beq 
V_1(t)=V_a+V_b\sin{({2\pi\over \tau} t)} \; , 
\eeq 
where $\tau$ is the period of oscillation. 
\par
The NPSE is integrated by using a finite-difference 
predictor-corrector method [13,14,15]. 
The initial wave function of the condensate is found 
by solving the equation with imaginary time.  
In Figure 1 we plot the axial density profile $\rho(z)$ 
of the matter-waves coming out from the potential well 
with $\tau=1.5$. The figure shows that, as expected, 
the emission of pulses is periodic. Moreover, 
both the period of emission and the tunneling 
fraction $P_T$ depend on $\tau$. 
In the inset of Figure 2 the tunneling probability $P_T$ is shown 
as a function of $\tau$ for a Bose condensate 
with $N=10^3$ and $N=10^4$ atoms. 
The emission probability has its absolute maximum 
near $\tau=0.8$, that is the average period 
of the oscillations of the center of mass 
of the condensate inside the potential well. 
\par 
The enhancement of the emission probability of coherent 
matter near $\tau=0.8$ has a classical explanation. 
We have verified [6] that it corresponds to the condition of resonance 
between the oscillation of the energy barrier and 
the oscillation of the center of mass of the condensate. 
Close to the resonance condition the phenomenon of coherent emission 
of atoms is no more due to quantum tunneling but to the parametric 
resonance [16] between the period of 
oscillation of the center of mass of the condensate 
and the period of oscillation of the energy barrier. 
This might explain the different $N$ dependence of $P_T$ and $P_T^0$ 
as shown in Figure 2: $P_T^0$ is dominated by quantum effects and, as such, 
is enhanced by interaction, while $P_T$ is essentially a classical 
effect and is almost $N$ independent, in agreement with the 
numerical data shown in Figure 2. 
\par 
Note that in Figure 2 there are other local maxima 
whose positions follow the text-book [16] resonant condition formula 
$\tau = \left(n/2\right) \tau_{osc}$, where $\tau_{osc}$ is the period 
of oscillation of the center of mass in the absence of an oscillating 
barrier, with $n$ an integer number. 

\section*{Conclusions} 

In this paper we have studied the controlled 
emission of coherent matter pulses from a trap by changing 
the period of oscillation of the energy barrier. 
The experimental investigation of the  
physical configuration we have considered in this paper 
may contribute to the realization of novel phenomena in atom lasers. 
Finally, we observe that the parametric driving of Bose-Einstein 
condensates can be obtained by current experiments using optical dipole 
forces with far-detuned laser beams. 
By varying the intensity of the laser beams one can control 
the height of the energy barrier that confines the condensed sample. 

\section*{References}

\begin{description}

\item{\ [1]} S. Fribel, C. D'Andrea, J. Walz, M. Weitz, 
and T.W. H\"ansch, Phys. Rev. A {\bf 57} R20 (1998).   

\item{\ [2]} R. Jauregui, N. Poli, G. Roati, and G. Modugno, 
Phys. Rev. A {\bf 64} 033403 (2001).   

\item{\ [3]} P.G. Kevrekidis, A.R. Bishop, and K.O. Rasmussen, 
J. Low Temp. Phys. {\bf 120} 205 (2000). 

\item{\ [4]} V.R. Manfredi and L. Salasnich, Int. J. Mod. Phys. B 
{\bf 13}, 49 (1999); Manfredi and L. Salasnich, in A. Fabrocini et al. (Eds.), 
Perspectives on Theoretical Nuclear Physics VII, pp. 319-324 
(Edizioni ETS, Pisa, 1999). 

\item{\ [5]} L. Salasnich, Phys. Lett. A {\bf 266} 187 (2000); 
L. Salasnich, Progr. Theor. Phys. Suppl. {\bf 139} 414 (2000). 

\item{\ [6]} L. Salasnich, A. Parola, L. Reatto, 
J. Phys. B {\bf 35}, 3205 (2002).  

\item{\ [7]} L. Salasnich, Laser Physics {\bf 12} 198 (2002). 

\item{\ [8]} L. Salasnich, A. Parola, and L. Reatto,   
Phys. Rev. A {\bf 65} 043614 (2002). 

\item{\ [9]} E.P. Gross, Nuovo Cimento {\bf 20}, 454 (1961); 
L.P. Pitaevskii, Zh. Eksp. Teor. Fiz. {\bf 40}, 
646 (1961) [English Transl. Sov. Phys. JETP {\bf 13}, 451 (1961)]

\item{\ [10]} A.J. Leggett, Rev. Mod. Phys. {\bf 73}, 307 (2001). 

\item{\ [11]} L. Tonks, Phys. Rev. {\bf 50}, 955 (1936); 
M. Girardeau, J. Math. Phys. {\bf 1}, 516 (1960). 

\item{\ [12]} P. Massignan and M. Modugno, cond-mat/0205516; 
P. Massignan, Laurea Thesis, University of Milano, unpublished. 

\item{\ [13]} E. Cerboneschi, R. Mannella, E. Arimondo, and L. Salasnich, 
Phys. Lett. A {\bf 249}, 245 (1998). 

\item{\ [14]} L. Salasnich, A. Parola, and L. Reatto, Phys. Rev. A {\bf 59}, 
2990 (1999); L. Salasnich, A. Parola, and L. Reatto, Phys. Rev. A {\bf 60}, 
4171 (1999). 

\item{\ [15]} L. Salasnich, A. Parola, and L. Reatto, Phys. Rev. A {\bf 64}, 
023601 (2001). 

\item{\ [16]} Landau L D and Lifsits E M 1991 {\it Mechanics, 
Course of Theoretical Physics}, vol. 3 (Pergamon Press: Oxford); 
Arnold V I 1990 {\it Mathematical Methods of Classical Mechanics} 
(Springer: Berlin). 

\end{description}

\begin{figure}
\centerline{\psfig{file=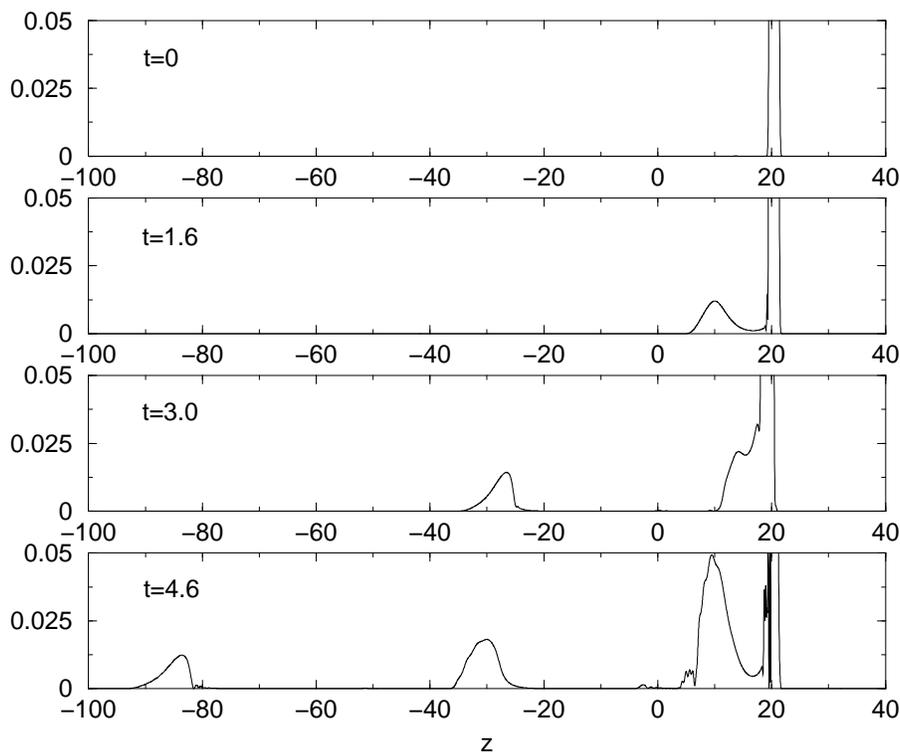,height=4.in}}
\caption{Axial density profile of Bose condensed $^{23}$Na atoms 
tunneling through the potential well 
(Eq. (13)) with oscillating barrier (Eq. (14)): 
$V_a=300$ and $V_b=200$. Results obtained by solving NPSE. 
Period of oscillation: $\tau =1.5$. 
Effective chemical potential of the 
initial condensate: $\mu_{eff}=57.43$. 
Number of $^{23}$Na atoms: $N=10^4$. 
Scattering length: $a_s=30$ $\AA$.  
Note that $P_T=0.26$ with $V_1(t)=V_a-V_b=100$. 
Length $z$ in units $a_z=(\hbar/m\omega_z)$, 
where $\omega_z=\omega_{\bot}/10$ with $\omega_{\bot}=2\pi$ kHz; 
energy in units $\hbar \omega_z$ and time in units $\omega_z^{-1}$.} 
\end{figure} 

\begin{figure}
\centerline{\psfig{file=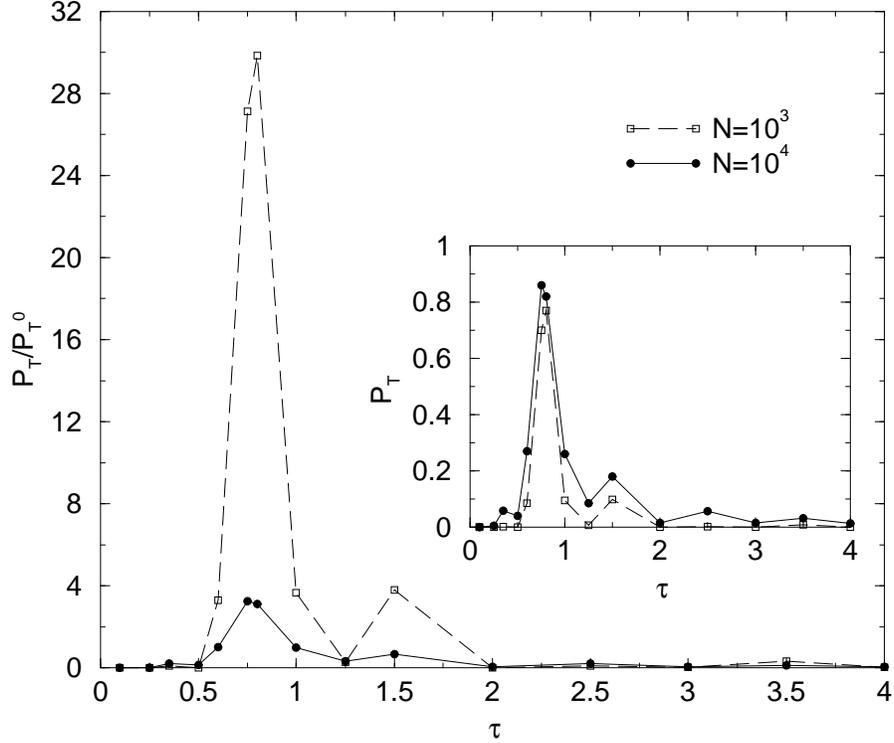,height=4.in}}
\caption{Tunneling ratio $P_T/P_T^0$ at $t=5$ as a function 
of the period $\tau$ of oscillation of the lower-lying 
Gaussian barrier (Eq. (13,14)), as obtained by solving NPSE,  
with $V_a=300$ and $V_b=200$ (Eq. (6)). 
The period $\tau_{osc}$ of small oscillations around the minimum of 
the potential well with $V_a=300$ and $V_b=0$ is $\tau_{osc}=0.84$. 
$P_T^0$ is the tunneling probability with $V_1=V_a-V_b=100$. 
Effective chemical potential of the 
initial condensate: $\mu_{eff}=27.15$ for $N=10^3$, 
$\mu_{eff}=57.43$ for $N=10^4$. 
$N$ is the number of $^{23}$Na atoms. 
Scattering length: $a_s=30$ $\AA$. Inset: tunneling probability 
$P_T$ at t=5 as a function of $\tau$. Units as in Fig. 1.} 
\end{figure} 
      
\end{document}